\documentclass[%
 reprint,
 amsmath,amssymb,
 aps,
]{revtex4-2}

\usepackage{graphicx}% Include figure files
\usepackage{dcolumn}% Align table columns on decimal point
\usepackage{bm}% bold math
\usepackage{enumitem}
\newcommand{\ellb}{\boldsymbol{\ell}}

\begin{document}

\preprint{APS/123-QED}

\title{Capillary Coarsening Drives Re-entrant Statistical Transitions in Early Coffee-Ring Growth}% Force line breaks with \\
%\thanks{A footnote to the article title}%

\author{Arundas Chandradasan}
 %\altaffiliation[Also at ]{Physics Department, XYZ University.}%Lines break automatically or can be forced with \\
\author{Dileep Mampallil}%
 \email{dileep.mampallil@iisertirupati.ac.in}
\affiliation{%
Indian Institute of Science Education and Research Tirupati, Yerpedu, 517619, AP, India
}%

%\collaboration{MUSO Collaboration}%\noaffiliation

\date{\today}% It is always \today, today,
             %  but any date may be explicitly specified

\begin{abstract}
Understanding how nonequilibrium systems evolve between distinct statistical regimes remains a fundamental challenge in statistical physics. Here we show that particle deposition at an evaporating contact line provides an experimental realization of cooperative one-dimensional deposition, in which post-deposition capillary attraction drives particles to migrate, cluster, and coarsen. As deposition proceeds, the cluster-width distribution undergoes a striking re-entrant transition, evolving from an initial exponential form to a peaked Gamma distribution and subsequently returning to an exponential form. This dynamic transition is accompanied by a non-monotonic crossover in fluctuation scaling, \(\sigma_w\sim\langle w\rangle^\kappa\), with \(\kappa\) changing successively from \(1\) to \(1/2\) and back to \(1\). Monte Carlo simulations reproduce these statistical transitions. Our results establish evaporative contact-line deposition as a simple model of correlated one-dimensional assembly and reveal how local interactions can produce re-entrant statistical behaviour in a nonequilibrium system.

\end{abstract}

%\keywords{Suggested keywords}%Use showkeys class option if keyword
                              %display desired
\maketitle

%\tableofcontents

\section{Introduction}

Evaporating drops are rich physical micro-laboratories, where hydrodynamic flows, self-assembly and complex deposit statistics emerge from a simple drying process \cite{deegan1997capillary, DeeganPRE, SEFIANE2014372, MampallilEral}. When particles accumulate near the contact line \cite{annurev-fluid-011212, GelderblomSM2022, marin2011order}, their collective dynamics provide a simple setting to study nonequilibrium fluctuations. A useful measure is Taylor’s fluctuation-scaling law, \(\sigma_X\sim\langle X\rangle^\kappa\), with \(\kappa\) characterizing different statistical regimes of the evolution of $X$ \cite{eisler2008fluctuation}. Independent particle arrivals give Poissonian growth fluctuations of the deposit with \(\kappa=1/2\), while correlated interfacial growth can exhibit different scaling, such as \(\kappa=1/3\) in the Kardar–Parisi–Zhang (KPZ) universality class \cite{PRL56.889,PRL106.250603,PRL108.190603,yunker2013effects,PRL209601Comment,Oliveira_2014, dias2018interaction}.

While radial interface growth has been extensively explored within these frameworks, the early-stage lateral assembly of particles along the outermost contact line presents a simpler, distinctly one-dimensional problem, closely related to random sequential adsorption (RSA) \cite{evans1993random, talbot2000car, kubala2022random}. In this classic Rényi's “car-parking problem”, hard particles arrive randomly and irreversibly occupy available space, approaching the jamming limit (\(\simeq0.7476\)) \cite{Renyi1963}. But what happens in a more realistic parking process, where a good driver finding a small gap moves towards a neighbouring car? Such cooperative RSA models with post-deposition diffusion have been studied theoretically \cite{evans1993random,budinski2001random,lavalle1999extended,PRL111.028302, PRL.122.130602} and experimentally with car parking itself \cite{rawal2005modelling}.  However, only close experimental realizations were achieved in microscale as spontaneous processes, such as the slow relaxation of adsorbed colloids at water/oil interfaces \cite{kaz2012physical,rahmani2016colloidal}. A microscale realization of this “realistic car-parking problem” is challenging, as it requires both one-dimensional deposition and strong, correlated post-deposition rearrangements.

Here we show that particle deposition at evaporating contact lines provides an experimental realization of such cooperative deposition. Rather than accumulating independently, deposited particles attract neighbouring particles, forming clusters that merge and coarsen along the contact line (Fig.~1A). Tracking the evolution of the cluster-width distribution reveals a striking re-entrant transition: it evolves from an initial exponential form to a peaked Gamma distribution and then returns to an exponential form (Fig.~1B). This evolution is accompanied by a nontrivial crossover in fluctuation scaling, \(\sigma_w\sim\langle w\rangle^\kappa\), with \(\kappa\) changing successively from \(1\) to \(1/2\) and back to \(1\) (Fig.~1C).

Although nonequilibrium systems with competing interactions can undergo transitions between distinct statistical regimes, including changes in fluctuation statistics and distributional form \cite{PRL040605,druagulescu2001exponential, PhysRevE.79.031117, lee2023size, becton2024dynamic, rouwhorst2020nonequilibrium}, such transitions are generally directional.  Re-entrant statistical behaviour, in which a system returns to its initial distributional form after passing through an intermediate state, has been predicted \cite{di2024brownian} but, to the best of our knowledge, never observed experimentally. The coffee-ring system therefore provides not only an experimental realization of correlated one-dimensional deposition, but also a simple setting in which to uncover unexpected statistical transitions. Our Monte Carlo simulations reproduce the experimental observations and show that capillary coarsening is sufficient to generate the successive statistical regimes, linking evaporative deposition to a broader class of nonequilibrium coarsening processes.

\begin{center}
   \begin{figure*}[ht] 
		\centering
		\includegraphics[scale= 0.96]{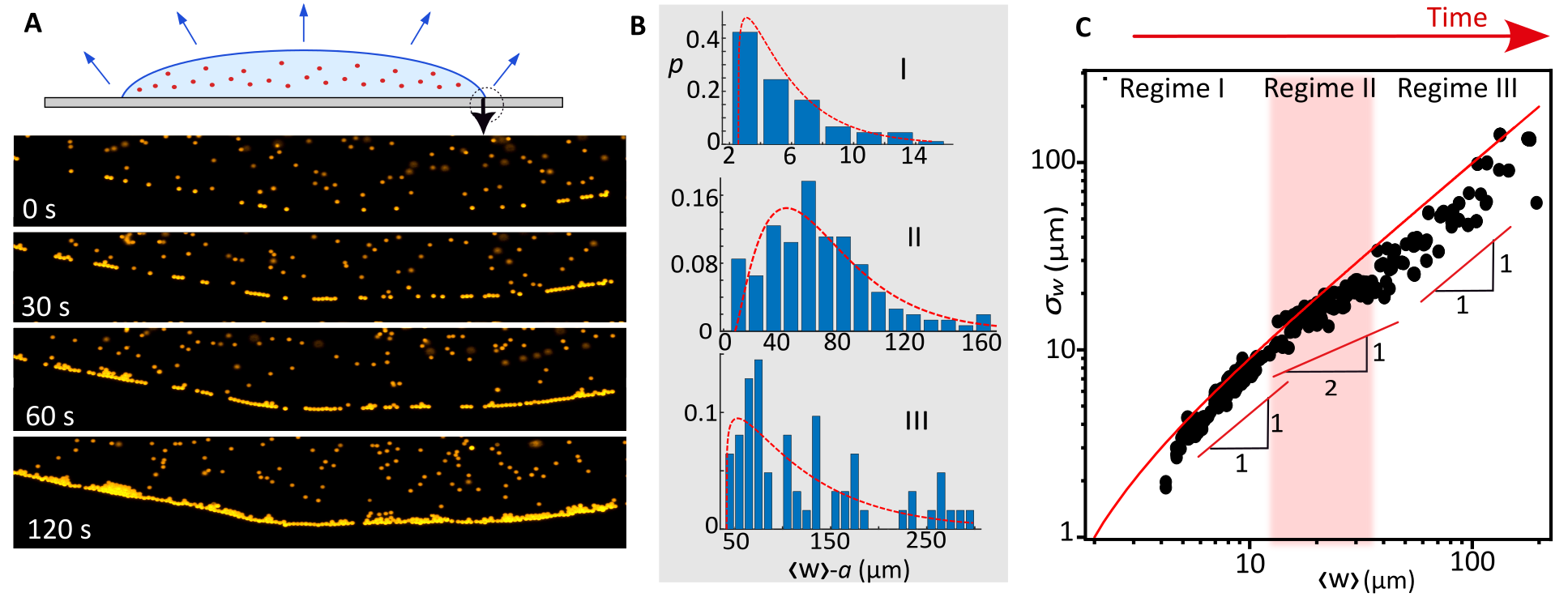}
		\caption{ (a) Time sequence of the contact-line region of an evaporating drop showing the growth of particle deposits through cluster formation. Neighboring clusters subsequently merge, producing a continuous ring well before complete evaporation of the drop. (b) Representative cluster-width distributions at the three regimes. Dashed curves are fits to the Gamma distribution (see main text). Data from three different experiments. (c) Evolution of the cluster-width distribution during ring growth, revealing three distinct statistical regimes characterized by the fluctuation scaling, $\sigma_w \sim \langle w\rangle^\kappa$, and the corresponding distribution shape: exponential-like, peaked, and exponential-like. The measured scaling exponents, $\kappa$, are indicated. The solid curve is $\sigma_w=\sqrt{\langle w \rangle}\sqrt{\langle w\rangle-1}$ (see main text). Particle diameter: $2~\mu$m. } \label{Fig1}
	\end{figure*}
\end{center}

\section{Results and discussion}

Immediately after the drop is placed on the substrate, the suspended particles start to arrive at the contact line.  We can observe that they form small linear clusters stretched along the rim (Fig.~1a). Finally, these clusters merge to form a full ring. 

When particles reach the liquid wedge near the contact line, they become partially exposed to air, creating curved liquid menisci around them. If two particles come within a capture distance, the overlap of their menisci generates a lateral capillary force \cite{kralchevsky1994capillary, paunov1993lateral, kralchevsky2000capillary}, $F_c= -2\pi\gamma Q^2/\ell$ where $\ell$ is the separation between the particles, $\gamma$ is the surface tension, $Q=r \sin\psi$ with $r$ is the contact line radius on the particles and $\psi$ is the meniscus angle with the horizontal. $\psi >0$ for hydrophilic particles and $\psi<0$ for hydrophobic particles. 

The drag force on a partially immersed particle \cite{hunt2023drag, zhou2022drag, dorr2016drag} can be written as $F_d = -f(\theta)\eta a\,(d\ell/dt)$, where $f(\theta)$ is a factor accounting for the partial immersion, $\theta$ is the contact angle, $\eta$ is the dynamic liquid viscosity, and $a$ is the particle diameter. In the low Reynolds number case, $\mathbf{F_c}+\mathbf{F_d}=\textbf{0}$,  can be simplified as 

\begin{equation}\label{eq_mot}
    \bar{\ell}\,^2=1-\bar{t}
\end{equation}

where $\bar{\ell} = \ell/\ell_0$ and $\bar{t}=t/t_f$, with $\ell_0$ being the initial particle separation and $t_f$ being the time taken for the particles to meet from the separation $\ell_0$ (see Methods).

\begin{center}
   \begin{figure}[ht] 
		\centering
		\includegraphics[scale= 0.96]{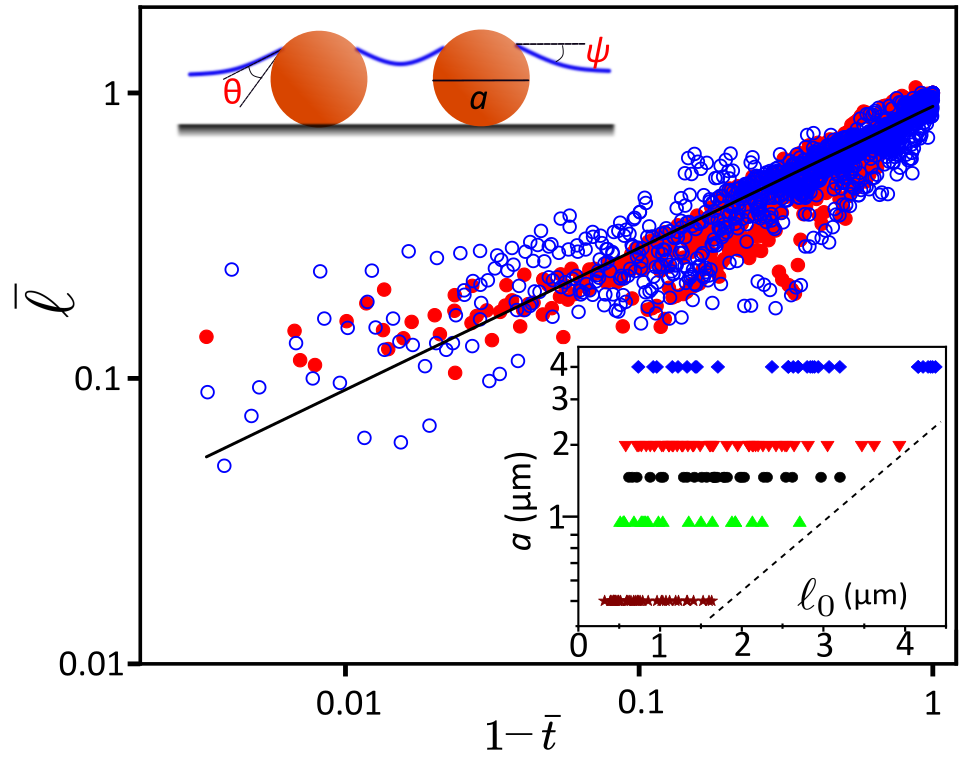}
		\caption{(a)  Normalized particle separation, $\bar{\ell}$, as a function of $1-\bar{t}$ for polystyrene particles of two different diameters. The solid line has a slope of $1/2$, consistent with Eq.~\ref{eq_mot}. (Inset): By observing multiple lateral migrations for particles of different diameters, we can estimate the capillary attraction range $\ell_c$ as the largest observed $\ell_0$.  The dashed line indicates the approximate linear scaling \(\ell_c\propto a\).  } \label{Fig2}
	\end{figure}
\end{center}

In Fig.~2(a), we plot $\bar{\ell}$ versus $1-\bar{t}$ for polystyrene particles of two different diameters (2 and 4~$\mu$m). For each lateral migration event, we normalized the data by the corresponding initial separation $\ell_0$ and migration time $t_f$. The data collapse onto the predicted scaling, consistent with Eq.~\ref{eq_mot}, confirming that lateral capillary attraction, rather than random deposition, drives the observed clustering. 

We have observed that the addition of surfactant decreases the attraction (Supplementary Fig.S1). 
On the other hand, the capillary clustering is slightly stronger in pendent drops, likely due to the absence of particle-substrate friction. Since the Bond number, $Bo=\Delta\rho g a^2/\gamma\ll1$ ( with $\Delta\rho$ being the particle-liquid density difference and $g$ being the acceleration due to gravity), particle weight does not significantly deform the interface, ruling out flotation capillary forces.

Deposition at the edge is a purely 1D system because the radially inward cluster growth does not affect the lateral width.  The probability of capturing a newly arriving particle at the contact line is determined by the meniscus deformation near the cluster tip. Thus, the growth is largely independent of the cluster size. Consequently, larger clusters do not possess a systematic advantage over smaller ones. This contrasts with the radial interface growth of ellipsoidal particle deposits, where anisotropic capillary interactions lead to preferential growth of protrusions and morphological instabilities \cite{yunker2013effects, PRL209601Comment, PPRL209602Reply}.

\begin{center}
   \begin{figure*}[ht] 
		\centering
		\includegraphics[scale= 0.99]{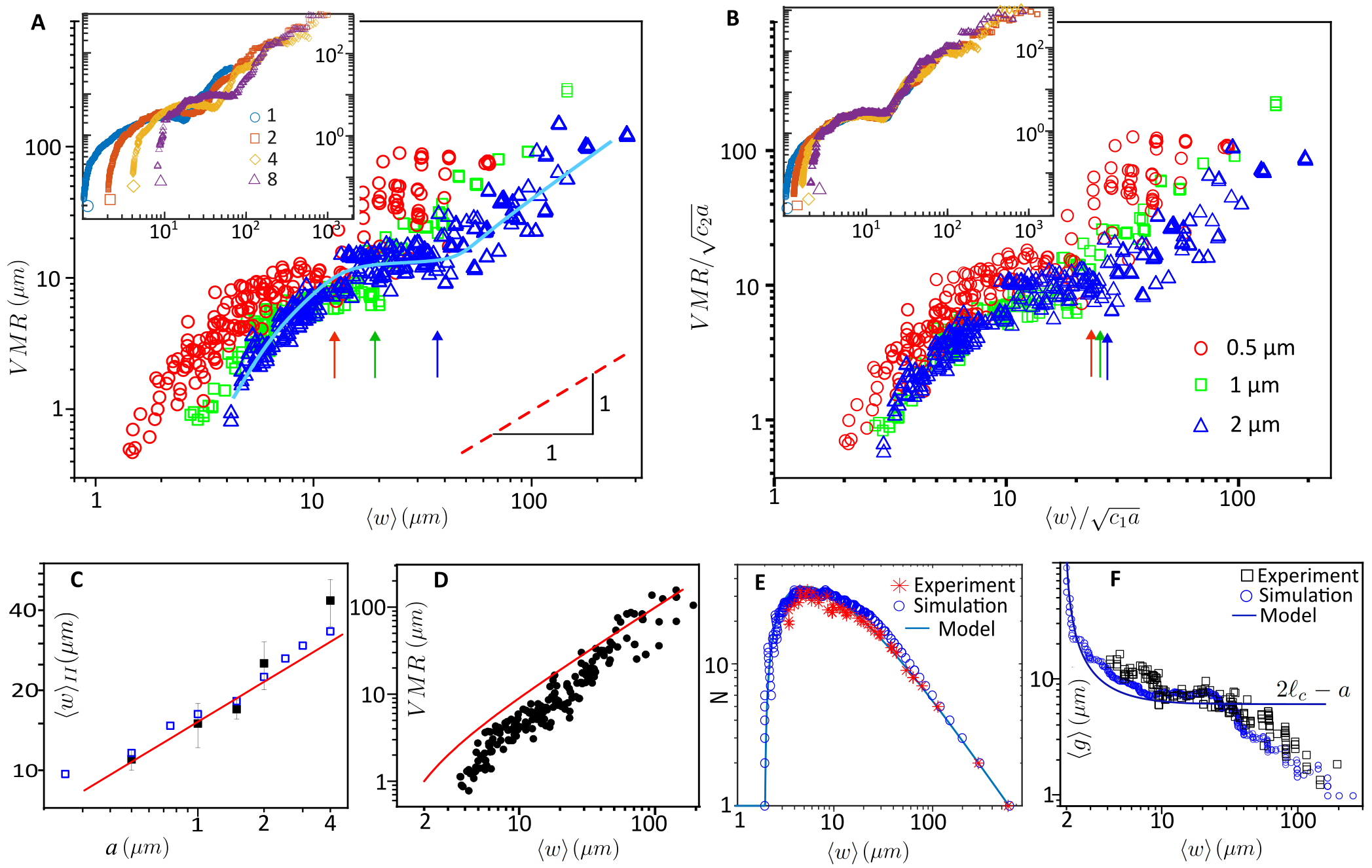}
		\caption{ \textbf{A} $\text{\small VMR}$ for particles of different diameters (0.5, 1, and 2 $\mu$m). The arrows indicate the cross-over points between Regimes II \& III ($\equiv \langle w\rangle_{II}$). The solid line is a guide to the eye. \textbf{B}. Rescaling $\langle w \rangle$ and $\text{VMR}$ by $\sqrt{a}$ collapses the data in Regime II onto a single crossover point. $c_1$ and $c_2$ ($1\,\mu\text{m}$) are dimensional prefactors ensuring dimensionless plot axes.  The inset of \textbf{A} and \textbf{B} shows Monte Carlo-simulated data for different particle units. \textbf{C}. The cross-over points $\langle w\rangle_{II}$ for different sized particles from experiments ($\blacksquare$) and simulation ($\square$). The solid line has a slope of $1/2$. \textbf{D}. When capillary clustering is weak, $\text{\small VMR}$ does not show Regimes II \& III.  The solid line is $\text{\small VMR}=\langle w\rangle-1$, obtained from the geometric distribution. \textbf{E}. Number of clusters versus $\langle w\rangle$. Data from four different experiments. \textbf{F}. Mean gap versus mean width from experiments ($\square$) and simulation ($\circ$). Shorter $\langle w\rangle$ for $t \rightarrow 0$ are not experimentally reachable. $a=2~\mu$m. With $\ell_c \approx 2 a$, the solid line is obtained from the mean-field model. Below $\langle g\rangle = 2\ell_c-a$, cluster births tend to zero, initiating Regime III. } \label{Fig3}
	\end{figure*}
\end{center}

\medskip
\noindent \textit{Ring growth statistics:} Having established the growth mechanism, we now examine the statistics of cluster growth in detail. The cluster width ranges from a single particle diameter to the ring perimeter. The width distributions evolve from an exponential-like form at early times to a peaked distribution at intermediate times, before reverting to an exponential-like form as lateral filling is completed (Fig.\ref{Fig1}B).

We further characterize the growth using the variance-to-mean ratio, $\text{\small VMR}=\sigma_w^2/\langle w\rangle$. Since $\sigma_w\sim\langle w\rangle^\kappa$, we expect $\text{\small VMR}\sim\langle w\rangle$ for $\kappa=1$ and a constant $\text{\small VMR}$ for $\kappa=1/2$. As shown in Fig.~3(a), the measured $\text{\small VMR}$  exhibits three successive scaling regimes with slopes of $1$, $0$, and $1$, corresponding to regimes I ($\kappa \rightarrow 1$), II ($\kappa=1/2$), and III ($\kappa=1$), respectively. 

In Fig.~3(a), we plot the VMR curves for different particle sizes. Since particles constitute the elementary units of the clusters, we expect \(\langle w\rangle\propto a\) for simple stochastic deposition. Interestingly, this scaling changes in Regime II. Here, because \(\sigma_w\propto\sqrt{\langle w\rangle}\) in this regime, we expect that the characteristic cluster width scales as \(\langle w\rangle\propto\sqrt{a}\). Accordingly, rescaling both the VMR and \(\langle w\rangle\) by a quantity proportional to \(\sqrt{a}\) collapses the curves for different particle diameters in Regime II. This scaling is particularly evident from the crossover between Regimes II and III, indicated by the arrows. The simulations reproduce the same Regime-II scaling, as shown in the insets of Fig.~\ref{Fig3}A\&B. Figure~\ref{Fig3}C summarizes this scaling over a range of particle sizes for both experiments and simulations.

The VMR curves do not collapse in Regime I under the \(\sqrt{a}\) scaling. Instead, rescaling the mean cluster width as \(\langle w\rangle/a\) collapses the curves in Regime I, consistent with \(\langle w\rangle\propto a\) (Supplementary Fig.S2). This contrasting scaling indicates a change in the underlying aggregation dynamics between Regimes I and II. The apparent collapse in Regime III follows from the \(\sqrt{a}\) scaling established in Regime II, although the underlying physical processes in the two regimes are distinct.

When capillary clustering is negligible, the plateau in the VMR is absent (Fig.~3D). Although eliminating capillary attraction experimentally is difficult, it can be effectively suppressed by using particles of different sizes. In this case, smaller particles reside beneath the meniscus of larger particles and therefore experience negligible lateral capillary attraction, while hindering the larger particle's motion. Adding surfactants also decreases the capillary clustering effect. Consistent with the experiments, our simulations reproduce the presence or absence of Regime II depending on whether capillary clustering is switched on or off (Supplementary Fig.S3).

\medskip

\textit{Mechanism behind the statistical transitions}: Let us now look at the deposition process and the evolution of the observed scalings in different temporal regimes. 

The particle arrival is a Poisson process, and the cluster width corresponds to the waiting distance between successive deposition events. Therefore, the width distribution must follow geometric statistics when particle-particle interactions are negligible,  yielding an exponentially decaying distribution. In the discrete description, the cluster-width distribution is $P(w)=(1-p)p^{w-1}$, where $p$ is the deposition probability. In the continuum limit, this reduces to $P(w)\sim\exp(-w/w_0)$, with $w_0=-1/\ln p$. The corresponding mean is $\langle w\rangle=1/(1-p)$ and variance is $\mathrm{Var}(w)=p/(1-p)^2$.

At early times, i.e., in regime I, capillary clustering is absent as the average gap between the particles $\langle g\rangle \gg \ell_c$. Therefore, cluster widths follow a  geometric distribution giving  $\sigma_w=\sqrt{\langle w\rangle}\sqrt{\langle w\rangle-1}$ and $\text{\small VMR} = \langle w\rangle-1$. Consequently, for $\langle w\rangle\gg1$, the scaling approaches $\sigma_w\sim\langle w\rangle$, consistent with the end of regime I in Figs.~1C and Fig.3A.
\medskip

In the intermediate regime II, Capillary coarsening progressively eliminates smaller clusters through repeated merging, producing a pronounced peak in the width distribution histogram (Figs.~1B). Since $\langle g\rangle > \ell_c$ (see Fig.\ref{Fig3}E), spatial independence of these merging processes dictates central limit behavior ($\operatorname{Var}(w) \propto \langle w \rangle$), yielding a constant $\text{\small VMR}$.

In Regime III, smaller clusters progressively coalesce into larger macro-clusters, depleting the initial peak in the cluster-size distribution. The system becomes dominated by wide clusters separated by narrow gaps, with ($\langle g\rangle < 2\ell_c$). Particles arriving within these gaps are immediately captured by the neighbouring clusters rather than nucleating as new clusters. This causes the neighbouring clusters to grow towards each other and eventually merge. Particle arrivals into these isolated gaps become effectively Poissonian, with exponentially distributed waiting times. This memoryless kinetics restores the exponential cluster-width distribution observed in Regime I. Thus, the loss of self-averaging brings back \(\operatorname{Var}(w)\propto\langle w\rangle^2\) and hence \(\sigma_w\propto\langle w\rangle\), leading to the linear growth of the VMR.
\medskip

In short,
\begin{enumerate}[label=\roman*)]
\item In the capillary-clustering process, two subtly different mechanisms govern lateral cluster growth: \textit{lateral sliding and subsequent coalescence} of neighbouring clusters in Regime II, followed by the \textit{growth of adjacent clusters towards each other} and their eventual merger in Regime III.

\item Regimes I and III are governed by different stochastic processes, namely, the run length of cluster growth and the waiting time for gap closure, respectively. Yet, their analogous memoryless statistics give rise to the same exponential cluster-width distribution.
\end{enumerate}

To establish the theoretical mechanism driving this transition, we formulate a framework grounded in the Law of Total Variance. Conditioning the spatial width variance $\operatorname{Var}(w)$ on the constituent merger count $n$ yields (see Supplementary Information):

\begin{equation}
\begin{split}
\operatorname{Var}(w) &= \mathbb{E}[\operatorname{Var}(w \mid n)] + \operatorname{Var}(\mathbb{E}[w \mid n]) \\
&= \langle n \rangle \sigma_0^2 + \mu_0^2 \operatorname{Var}(n)
\end{split}
\end{equation}

where $\mu_0$ and $\sigma_0^2$ are the spatial mean width and variance of an elementary unmerged sub-cluster, respectively. This variance decomposition proves that the transition from Regime II to Regime III reflects a fundamental breakdown of additive self-averaging. 

In Regime II, spatial self-averaging across wide gaps bounds merger history fluctuations ($\operatorname{Var}(n) \propto \langle n \rangle$), leaving the linear within-group term $\mathbb{E}[\operatorname{Var}(w \mid n)] \propto \langle w \rangle$ dominant ($\sigma_w \sim \langle w \rangle^{1/2}$). In Regime III, the collapse of available gaps destroys spatial self-averaging, inducing extreme structural heterogeneity across macro-cluster growth histories ($\operatorname{Var}(n) \propto \langle n \rangle^2$). The quadratic structural term $\operatorname{Var}(\mathbb{E}[w \mid n]) \propto \langle w \rangle^2$ completely overwhelms micro-scale self-averaging, driving the re-entrant shift back to scale-invariant exponential statistics.

\medskip

\textit{A mean-field view}: Addressing the temporal regimes of the clustering process using a mean-field approach will give us some insights into the number evolution of the clusters and quasi-steady intermediate limits of the width and gaps. 

Along a 1D interface of length $L$, at deposition flux $J$, mass accumulates on available empty space $E(t) = L - M(t)$ at a rate $\mathrm{d}M/\mathrm{d}t = a J E(t)$. The net rate of evolution of the cluster population is determined by the balance between independent cluster birth ($R_b$) and capillary-induced coalescence ($R_c$) as $dN(t)/dt=R_b-R_c$ (see Methods for detailed model).
 
Cluster birth requires an arriving particle to land outside the capillary attraction zones of existing clusters ($g > 2\ell_c$). Defining the dimensionless interaction parameter $\chi(t) \equiv \ell_c / \langle g \rangle(t)$, the birth rate is given by $R_b = J E e^{-2\chi}$. Conversely, when gap spacings fall within the interaction range ($g \le 2\ell_c$), overlapping capillary capture zones trigger lateral coalescence at a rate $R_c = J E (e^{-2\chi} + 2\chi - 1)$. The net rate equation simplifies to

\begin{equation}
    \frac{\mathrm{d}N}{\mathrm{d}t} =  J E(t) - 2 \ell_c J N(t).
\end{equation}

As plotted in Fig.\ref{Fig3}E, integrating this population balance predicts mean-field continuous $N(t)$, capturing the trends observed in the experiments and simulation. We observe that the number of clusters sharply increases and becomes roughly stable in Regime II ($R_b \approx R_c$). In this self-averaging regime, the newly arriving particles are redistributed via capillary merging without substantially increasing the number of clusters. Subsequently, $N$ decreases sharply when $R_b\ll R_c$ in Regime III.  

The model reveals that the crossover $R_b = R_c$ corresponds to $\langle g \rangle = 2\ell_c$. Also, the limit of Regime II corresponds to $\langle g \rangle = 2\ell_c-a$, beyond which, in Regime III, the cluster nucleation tends to shut down ($R_b \to 0$) (Methods). A crossover at this limit can be observed both in the experiments and simulations (Fig.~\ref{Fig3}F), indicating the boundary between Regimes II \& III.   While mean-field theory captures limiting bounds, it misses Regime III, where growth shifts from additive coarsening to boundary accretion. In Regime III, the mean gap ($\langle g \rangle = \sum g_i/N$) drops as particle attachment at cluster tips continuously narrows the gaps, whereas the cluster count $N$ drops only in discrete steps upon complete gap closure.  More rigorous approaches, similar to the theoretical framework developed in Ref.~\cite{PRL.122.130602}, may capture such transition in the dynamics analytically.

\medskip

\textit{Connecting the temporal regimes}: Finally, we can connect the gradual statistical transitions, \textit{exponential $\rightarrow$ peaked $\rightarrow$  exponential}, by considering a Gamma distribution since this distribution provides a natural description of positive-valued stochastic variables arising from cumulative random growth processes \cite{frank2009common, doeschl2011implications, chen2023comprehensive}. 

The Gamma distribution is given as $P(w) = w^{k-1}\beta^{-k} \exp(-w/\beta)/\Gamma(k)$, where $k$ is the shape parameter and $\beta$ is the scale parameter. When $k \rightarrow 1$, the distribution tends to be exponential, and $k>1$ gives a peaked distribution. The distribution also gives $\langle w\rangle=k\beta$ and $\mathrm{Var}(w)=k\beta^2$, resulting in the relation, 
\begin{equation}
    \sigma_w=\langle w\rangle/\sqrt{k}. \label{eq:sigma-k}
\end{equation} 

In Regime II,  we expect $k\approx \langle w\rangle >1$,  for $\sigma_w \propto \langle w\rangle^{1/2}$. In Regime III, $k\approx1$, bringing the scale to $\sigma_w \sim \langle w\rangle$.

\begin{center}
   \begin{figure}[ht] 
		\centering
		\includegraphics[scale= 0.96]{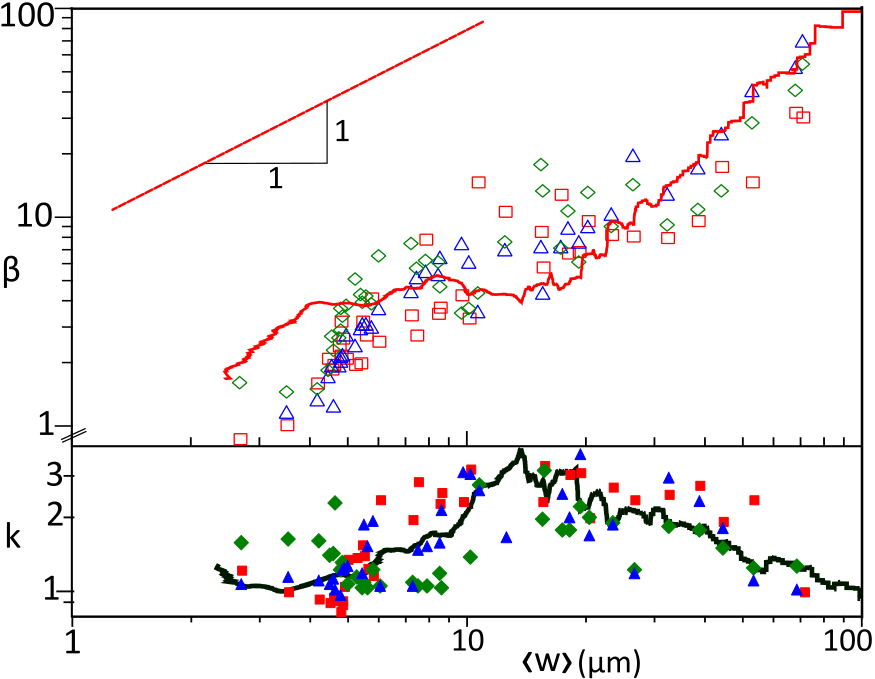}
		\caption{The shape parameter $k$ and scale parameter $\beta$ obtained by fitting the simulated (lines) and experimental (symbols) cluster widths at every time step with the Gamma distribution. Symbols represent four different experiments. $a=2~\mu$m. Since $\beta = \text{\small VMR}$, the $\beta$ data show the same characteristics as $\text{\small VMR}$ obtained directly from width data in Fig.3A.} \label{Fig4}
	\end{figure}
\end{center}

In Fig.~\ref{Fig1}B, Gamma distributions were fitted to the data in regimes I–III using maximum likelihood estimation, yielding shape parameters $k=1.1, 2.3$, and $1.2$, respectively. We applied the same analysis to the experimental and simulated width distributions throughout deposition (at every measured time step). In Fig.\ref{Fig4}, we can see that $k\approx1$ during the initial and final stages, consistent with an exponential distribution, while $k>1$ at the intermediate stage, indicating a pronounced peak. 

Interestingly, it can be seen that the scale parameter satisfies $\beta=\text{\small VMR}$. As shown in Fig.~\ref{Fig4}, the fitted values of $\beta$ reproduce the three statistical regimes, including the plateau in regime II. Thus, the variance-to-mean ratio obtained directly from the experimental data is equivalent to the scale parameter of the fitted distribution. This agreement demonstrates that the fluctuation transitions are intrinsically linked to the evolution of the underlying Gamma distribution, rather than representing an independent statistical measure.

\medskip

\textit{Discussion}: Exponential size distributions arise naturally in stochastic growth processes where individual growth events are statistically independent, and no characteristic size emerges. Competition between particle creation and removal through merging, mediated by local finite-range interactions within $\ell_c$ introduces a characteristic cluster size that evolves with deposition, producing a peaked, Gamma-like distribution. Similar peaked distributions occur in systems governed by correlated coalescence, including raindrop populations \cite{caracciolo2006analysis, baltas2016statistical, wang2025geographical}, nucleation--growth--coarsening systems \cite{Robson2004, zhc2-ggh3}, and diffusion-limited aggregation \cite{PhysRevLett.47.1400, PhysRevLett.57.2037, becton2024dynamic}. However, coalescence or diffusion-mediated merging does not necessarily generate a characteristic size: rapid, continuous nucleation followed by diffusion-driven random coalescence has been shown to preserve a self-similar exponential distribution throughout growth while slow nucleation shows power-law behaviour \cite{lee2023size}.

The re-entrant transitions reported here are fundamentally different from the re-entrant phase transitions in systems with competing interactions, such as liquid crystals \cite{PhysRevE.83.020701, PhysRevLett.86.3048} or liquid condensate phase of proteins \cite{krainer2021reentrant, pullara2022temperature}. Unlike conventional re-entrant thermodynamic phase transitions, the quantity that re-enters in this reported phenomenon is the statistical universality class of the fluctuations whose dynamic trajectory can be mapped through the parameter space of the Gamma distribution family. 

Interestingly, capillary clustering that we reported is indeed analogous to the "good driver's car parking problem", where local repositioning of parked cars produces Gamma-like distributions of inter-car gaps \cite{rawal2005modelling}. Similar long-tailed gap or length distributions have also been reported in cooperative protein binding along DNA \cite{van2008monte, lim2013nucleoid, kowalczykowski1986cooperative}. These parallels suggest that local interactions following stochastic deposition can generate nontrivial spatial statistics across systems ranging from everyday parking to biological macromolecules.

\section{Methods}
\subsection{Experiments}
Our experimental system involves water droplets in sessile and pendent forms, with typical volumes of 2 $\mu$l, containing polystyrene or silica particles, placed on glass slides and allowed to evaporate. The particle size ranged from 0.5 to 4 $\mu$m.  Tetradecyltrimethylammonium bromide (TTAB) was added when decreased surface tension was required. The particle accumulation at the drop edge was monitored using an inverted fluorescence microscope (Nikon Ti2) and camera (Nikon) until the edge was filled with particles without a gap. The gap width and particle cluster width were determined by a homemade Python code.  

\subsection{Monte Carlo simulations}

The contact line was represented by a linear domain of fixed length $L$, and particles were sequentially deposited at uniformly and randomly selected positions. In the absence of interactions, particles remained at their arrival locations, leading to stochastic cluster formation solely through random deposition. To incorporate capillary interactions, a newly arriving particle located within a distance $\ell_c$ from an existing cluster was irreversibly incorporated into that cluster. If multiple clusters were present within the capture range, the particle was attached to the nearest cluster. Inspired by the experimental observations, only clusters with mass (total number of particles in the cluster) less than a threshold value (typically 10) were allowed to move. This preserves the large clusters frozen as we observe in the experiments. Deposition continued until complete occupation of the contact line, and the resulting cluster-width distributions and fluctuation statistics were analyzed.

\subsection{Mathematical Methods}

\textbf{Capillary force calculation}: The capillary force between two particles at the liquid-air interface is $\mathbf{F_c}(\ell) = -2\pi\gamma Q_1Q_2 \,qK_1(q\ell)\, \hat{\ellb}$, where $\ell$ is the separation between the particles, $\gamma$ is the surface tension, $q = \sqrt{\rho g/\gamma}$, and $K_1$ is the modified Bessel function of first order. Higher-order terms are neglected as $q\ell \ll1$. Since the particles are of the same size and type, $Q_1=Q_2=r \sin\psi$, where $r$ is the contact line radius on the particles and $\psi$ is the meniscus angle with the horizontal such that $\psi >0$ for hydrophilic particles and $\psi<0$ for hydrophobic particles. We neglect any variation in $r$ due to the wedge shape of the drop edge. For $q\ell \ll1$ and $r\ll\ell$, we have $F_c= -2\pi\gamma Q^2/\ell$. 
\medskip

\textbf{Kinetic Model Derivation}: We consider a 1D contact line of length $L$. The rate of mass accumulation is $dM/dt = aJE(t)$, where $J$ is the arrival rate (particles per unit length per unit time) and $E$ is the total empty length. Integrating with $M(0)=0$, we get $M(t)$ and $E(t) = L-M(t)$. At $t=0$, the particle deposition at the contact line begins.

When $g>\ell_c$, the available gap length for cluster birth is $(g-2\ell_c)$. When $g<\ell_c$, the overlap length is $(2\ell_c-g)$. Assuming that the gap lengths follow a Poisson distribution $P(g) = \frac{1}{\langle g \rangle} e^{-g/\langle g \rangle}$, 
average uncaptured length per gap is 
 $$\langle l_{free}\rangle  = \int_{2\ell_c}^{\infty} (g - 2\ell_c) P(g) \, dg = \langle g\rangle e^{-2\chi}.$$  

The expected overlap length per gap is  
$$\langle l_{overlap} \rangle = \int_0^{2\ell_c} (2\ell_c - g) P(g) \,\mathrm{d}g = \langle g \rangle (e^{-2\chi} + 2\chi - 1),$$ 

where $\chi = \ell_c N / E$. With total $N$ gaps, we get $R_b=JN\langle l_{free}\rangle$ and total coalescence rate $R_c = J N \langle l_{overlap} \rangle$. Or, $R_b = J E e^{-2\chi}$ and $R_c = J E \left( e^{-2\chi} + 2\chi-1\right)$. 

Solving the linear differential equation, $dN/dt = R_b-R_c$, with initial condition $N(0) = 0$ gives the analytical cluster population:

   $$N(t) = \frac{L}{2\ell_c - a} \left( e^{-aJt} - e^{-2\ell_c Jt} \right).$$
   
The mean cluster width trajectory is obtained directly via $\langle w \rangle(t) = M(t)/N(t) = L\left(1 - e^{-aJt}\right) / N(t)$. Using $N(t)$ and $E(t)$, we can also obtain $\langle g\rangle (t) = (2\ell_c-a)\left( 1- e^{-(2\ell_c-a)Jt}\right)^{-1}$. For $Jt\gg1$, we get the limit $\langle g\rangle (t) \approx (2\ell_c-a)$. 

Our calculations use $J=1$, $a=2~\mu$m, $\ell_c=2a$, and $L=600~\mu$m, the length of the individual drop-edge segments imaged.

\bibliography{Reference}% Produces the bibliography via BibTeX.

\end{document}